\documentclass{jkas}

\def\year{2017} 
\def\month{June} 
\def\volume{50} 
\def\issue{3} 
\def\beginpage{999} 
\def\endpage{999} 
\def\received{May 4, 2017} 
\def\accepted{June 8, 2017} 
\date{Received \received; accepted \accepted}

\newcommand{\hii}{{\sc H~ii}\/ }

\def\simlt{\lower.5ex\hbox{$\; \buildrel < \over \sim \;$}}
\def\simgt{\lower.5ex\hbox{$\; \buildrel > \over \sim \;$}}
\def\arcdeg{\hbox{$^\circ$}}
\def\arcmin{\hbox{$^\prime$}}
\def\arcsec{\hbox{$^{\prime\prime}$}}

\title{
$BVI$ Photometric Study of the Old Open Cluster Ruprecht 6
}

\author[1,2,3]{Sang Chul KIM}
\author[1]{Jaemann Kyeong}
\author[1,2,3]{Hong Soo Park}
\author[1,4]{Ilseung Han}
\author[1,2]{Joon Hyeop Lee}
\author[5]{Dae-Sik Moon}
\author[1]{Youngdae Lee}
\author[1,4]{Seongjae Kim}

\affil[1]{Korea Astronomy and Space Science Institute, 776 Daedukdae-ro, 
  Yuseong-gu, Daejeon 34055, Republic of Korea; 
  \email{sckim, jman, hspark, jhl, ylee@kasi.re.kr}}
\affil[2]{Korea University of Science and Technology (UST), 217 
  Gajeong-ro, Yuseong-gu, Daejeon 34113, Republic of Korea}
\affil[3]{Visiting astronomer, Cerro Tololo Inter-American Observatory, National Optical Astronomy Observatory, which is operated by the Association of Universities for Research in Astronomy (AURA) under a cooperative agreement with the National Science Foundation.}
\affil[4]{Kyungpook National University, Republic of Korea;
  \email{ihan@knu.ac.kr, ksj6283@naver.com}}
\affil[5]{Department of Astronomy and Astrophysics, University of Toronto, Toronto, ON M5S 3H4, Canada;
  \email{moon@astro.utoronto.ca}}

\def\corrauthor{
Sang Chul KIM
}

\def\runningauthor{
S. C. KIM et al.
}

\def\runningtitle{
Old Open Cluster Ruprecht 6
}

\def\keywords{
open clusters and associations: individual (Ruprecht 6)  --
Galaxy: disk -- Galaxy: stellar content
-- Galaxy: structure -- Hertzsprung-Russell diagram -- CTIO:1.0m
}

\def\abstracttext{
We present a $BVI$ optical photometric study of the old open cluster Ruprecht 6
  using the data obtained with 
  the SMARTS 1.0 m telescope at the CTIO, Chile.
Its color-magnitude diagrams show the clear existence of the main-sequence stars,
  of which turn-off point is located around $V \approx 18.45$ mag and $B-V \approx 0.85$ mag.
Three red clump (RC) stars are identified at $V = 16.00$ mag, $I = 14.41$ mag and $B-V = 1.35$ mag.
From the mean $K_s$-band magnitude of RC stars ($K_s=12.39 \pm 0.21$ mag) 
  in Ruprecht 6 from 2MASS photometry and the known absolute magnitudes of the RC stars
  ($M_{K_S} = -1.595 \pm 0.025$ mag), 
  we obtain the distance modulus to Ruprecht 6 
  $(m-M)_0 = 13.84 \pm 0.21$ mag ($d=5.86 \pm 0.60$ kpc).
From the $(J-K_s)$ and $(B-V)$ colors of the RC stars,
  comparison of the $(B-V)$ and $(V-I)$ colors of the bright stars in Ruprecht 6
  with those of the intrinsic colors of dwarf and giant stars,
  and the PARSEC isochrone fittings, 
  we derive the reddening values of 
  $E(B-V) = 0.42$ mag and $E(V-I) = 0.60$ mag.
Using the PARSEC isochrone fittings onto the color-magnitude diagrams,
  we estimate the age and metallicity to be:
  log (t) $=9.50 \pm 0.10$ (t $=3.16 \pm 0.82$ Gyr)
  and [Fe/H] $= -0.42 \pm 0.04$ dex.
We present the Galactocentric radial metallicity gradient analysis
  for old (age $> 1$ Gyr) open clusters of Dias et al. catalog,
  which likely follow a single relation of 
  [Fe/H] $=(-0.034\pm0.007) R_{\mathrm{GC}} + (0.190\pm0.080)$ (rms = 0.201)
  for the whole radial range
  or dual relation of 
  [Fe/H] $=(-0.077\pm0.017) R_{\mathrm{GC}} + (0.609\pm0.161)$ (rms = 0.152) 
  and constant ([Fe/H] $\sim -0.3$ dex) value,
  inside and outside of $R_\mathrm{GC} \sim 12$ kpc, respectively.
The metallicity and Galactocentric radius ($13.28\pm 0.54$ kpc) of Ruprecht 6 
  obtained in this study seem to be consistent with both of the relations.
}

\begin{document}
\jkashead 

\section{Introduction\label{sec:intro}}

The Galactocentric radial metallicity gradient in the disk of the Milky Way (MW)
  has been studied for a long time
  (e.g., \citealt{twarog97, park99, chen03, tadr03, kim06, carraro07, paunzen10, netopil16, 
  cantat16, jacobson16, tissera16}),
  for which open clusters (OCs) have been under intense studies,
  alongside with Cepheids \citep{andrievsky04,cescutti07},
  globular clusters \citep{yong08}, 
  red clump (RC) stars \citep{onal16},
  planetary nebulae \citep{stanghe10},
  \hii regions \citep{rudolph06,fern17}, field stars \citep{xiang15}, etc.
The disk metallicity gradient can give information on the disk
  formation processes, star formation processes, and chemical evolution
  of the spiral galaxies \citep{tissera16}.
One notable recent result is that of 
  \citet{and17}, who have claimed the slopes of 
  the Galactic radial metallicity gradient to be
  $\simeq -0.066$ dex kpc$^{-1}$ for the age range of $1-4$ Gyr and
  $\simeq -0.03$ dex kpc$^{-1}$ for $6-10$ Gyr,
  using 418 red giant stars observed by CoRoT-APOGEE
  with $6 < R_{\mathrm{GC}}$ \simlt 13 kpc and $\mid Z_{gal} \mid < 0.3$ kpc
  (where $R_{\mathrm{GC}}$ is the Galactocentric distance and
  $Z_{gal}$ is the vertical distance from the Galactic plane).

OCs, especially the old ones, are a good laboratory for verifying
  stellar evolution theories, and they are also good tracers of 
  the star formation and evolutionary history of the Galactic disk.
While \citet{lynga87} published data for over 1200 OCs,
  the observed number of OCs including candidates has increased to 2167
  in the latest version (3.5; 2016 January 28) of 
  \citet{dias02} (hereafter DAML02) 
  catalog\footnote{http://www.wilton.unifei.edu.br/ocdb/}.
Among the 2167 objects in the DAML02 OC catalog,
  only 298 (13.8\%) have metallicity estimates \citep{oliveira13,krisci15}, while
  703 (32.4\%) have radial velocity estimates,
  2013 (92.9\%) have age estimates,
  2025 (93.4\%) have reddening estimates,
  2040 (94.1\%) have distance estimates, and
  2104 (97.1\%) have proper motion estimates.
\citet{khar13} presented MW object catalog 
  for 3006 objects including 2808 OCs, 147 globular clusters and 
  51 associations, almost complete up to 1.8 kpc from the Sun, 
  where only 386 out of 3006 have metallicity values.
Considering the estimated total number of MW OCs ($\sim 100,000$;
  \citet{piskunov06,tadr11}), more surveys and detailed photometric and
  spectroscopic studies on individual clusters are awaited.

Ruprecht 6 \citep{rup66} is an old OC, located in the constellation of Canis Major and 
  at very large distance ($\sim 13.28 \pm 0.54$ kpc; see Table 1 below) from the Galactic center.
It is one of the small-size and
  poor OCs, which means it contains not so many number of member stars.
This could be the reason why there have been few studies on the cluster.
\citet{hase08}, which seems the only previous observational study on this cluster, 
  have used 65-cm telescope, AP8 1024 px CCD, and $BVI$ filters
  at Gunma Astronomical Observatory (FoV = $10.3\arcmin \times 10.3\arcmin$)
  to study 36 old OCs, and included Ruprecht 6 in their study.
They obtained some physical parameters for Ruprecht 6 
  from Padova isochrone fitting : age = 3.2 Gyr (${\rm log} ~t=9.50$),
  $Z=0.008$ ([Fe/H] $= -0.41$), $E(V-I)=0.60$ mag, and $(m-M)_0=14.43$ ($d=7.7$ kpc),
  and these values are enlisted in the DAML02 catalog.
Some basic physical parameters are summarized in Table~\ref{tab:infotab}.
The last parameter in Table~\ref{tab:infotab} presents 
  the Trumpler class of Ruprecht 6 to be III 1 p, 
  which means, respectively, (i) Ruprecht 6 is detached and shows no noticeable concentration,
  (ii) most stars in the cluster are of nearly the same apparent brightness,
  and (iii) it is poor containing less than 50 stars \citep{tru30,dias02}.
In this study, using the $BVI$ optical imaging data obtained by using the CTIO 1.0 m telescope,
  we present the photometric analysis and physical parameters 
  of the OC Ruprecht 6.

Section 2 describes the observations and data reduction processes.
Section 3 presents the results in this study: cluster center position,
  color-magnitude diagrams, reddening and distance estimations, and
  PARSEC isochrone fitting results.
In Sections 4 and 5, we discuss the Galactocentric metallicity distribution
  and summarize our results, respectively.

\begin{table*}[t!]
\caption{Basic Information of the Open Cluster Ruprecht 6\label{tab:infotab}}
\centering
\setlength{\tabcolsep}{1.2mm}
\begin{tabular}{lcl}
\toprule
Parameter & Ruprecht 6 & Reference \\
\midrule
Other name & C 0653-132 & SIMBAD \\
$\delta_{J2000}$, $\alpha_{J2000}$ &
  06$^h$ 56$^m$ 06$^s$, $-13$\arcdeg~ 15\arcmin~ 00\arcsec & \citet{hase08}\\
$l$,$b$ & 225.\arcdeg28, $-4.\arcdeg98$              & \citet{hase08}\\
Reddening, $E(B-V)$               & $0.42$ mag  & This study \\
Reddening, $E(V-I)$               & $0.60$ mag & This study \\
Distance modulus, $(m-M)_0$     & $13.84 \pm 0.21$ mag   & This study \\
Distance, $d$                     & $5.86 \pm 0.60$ kpc & This study \\
Galactocentric distance, $R_{\mathrm{GC}}$ & $13.28\pm 0.54$ kpc   & This study \\
Metallicity, [Fe/H]              & $-0.42 \pm 0.04$ dex & This study \\
Age, $t$                          & $3.16 \pm 0.82$ Gyr (${\rm log} ~t=9.50 \pm 0.10$) & This study \\
Trumpler type                     & III 1 p              & \citet{dias02}\\
\bottomrule
\end{tabular}
\end{table*}

\section{Observations and Data Reduction\label{sec:obsred}}

Observations have been done using the Small and Moderate Aperture Research Telescope
  System (SMARTS) 1.0 m telescope and Y4KCam CCD at the Cerro-Tololo Inter-American 
  Observatory (CTIO, Chile) on 2010 December 13 (UT).
The Y4KCam CCD made at the Ohio State University is composed of
  $4104 \times 4104$ pixels (each 15 $\mu$m, $\sim 0.289\arcsec$ pixel$^{-1}$) 
  at the Cassegrain focus of the telescope,
  while it becomes $4064 \times 4064$ pixels ($19.57\arcmin \times 19.57\arcmin$)
  after excluding the 40 pixel wide gaps on either horizontal and vertical center
  separating the CCD into four quadrants and reading out the pixel values
  \footnote{http://www.astronomy.ohio-state.edu/Y4KCam/detector}.

\begin{figure}
\centering
\includegraphics[width=80mm]{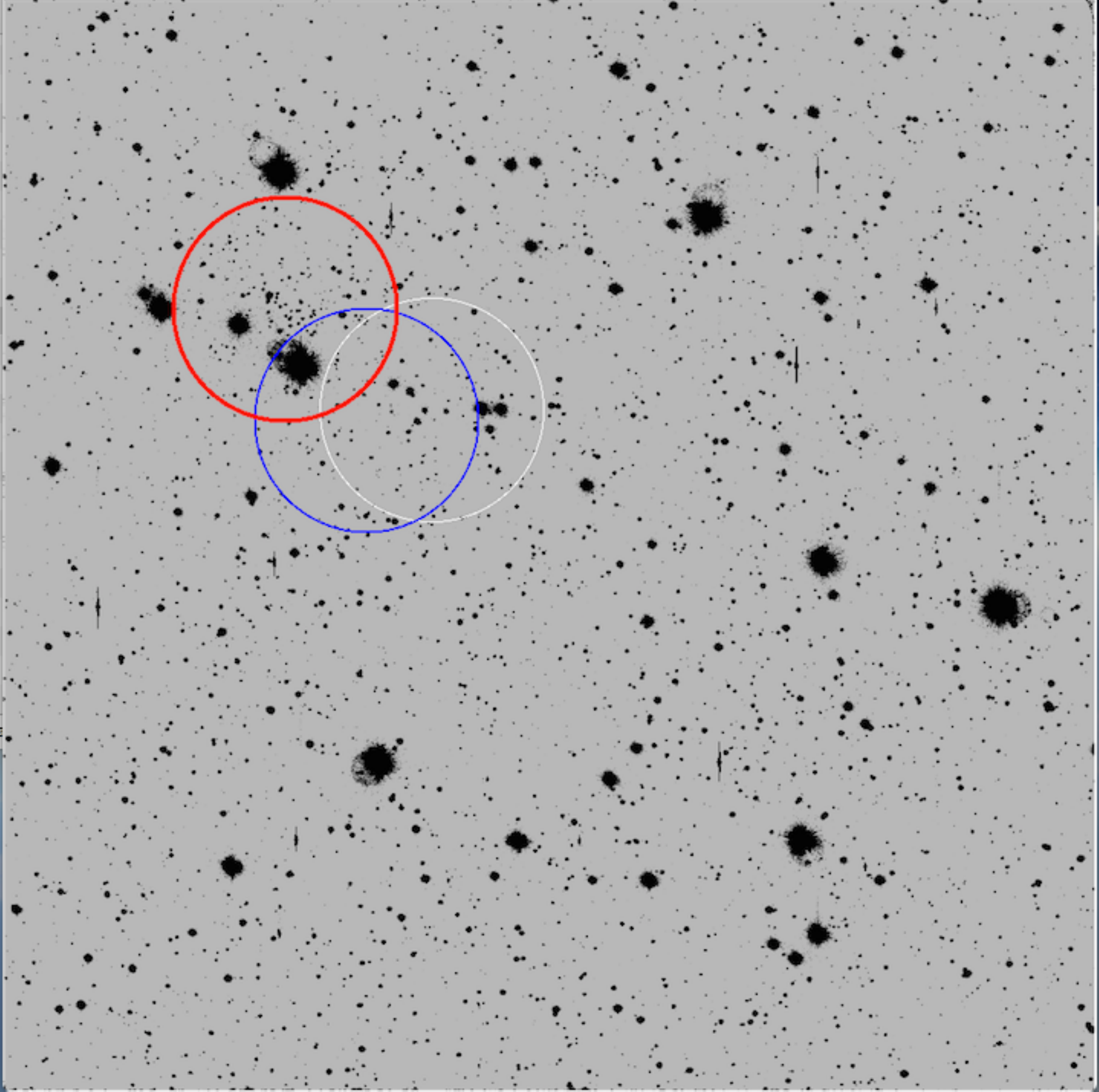}
\vskip 1mm
\includegraphics[width=80mm]{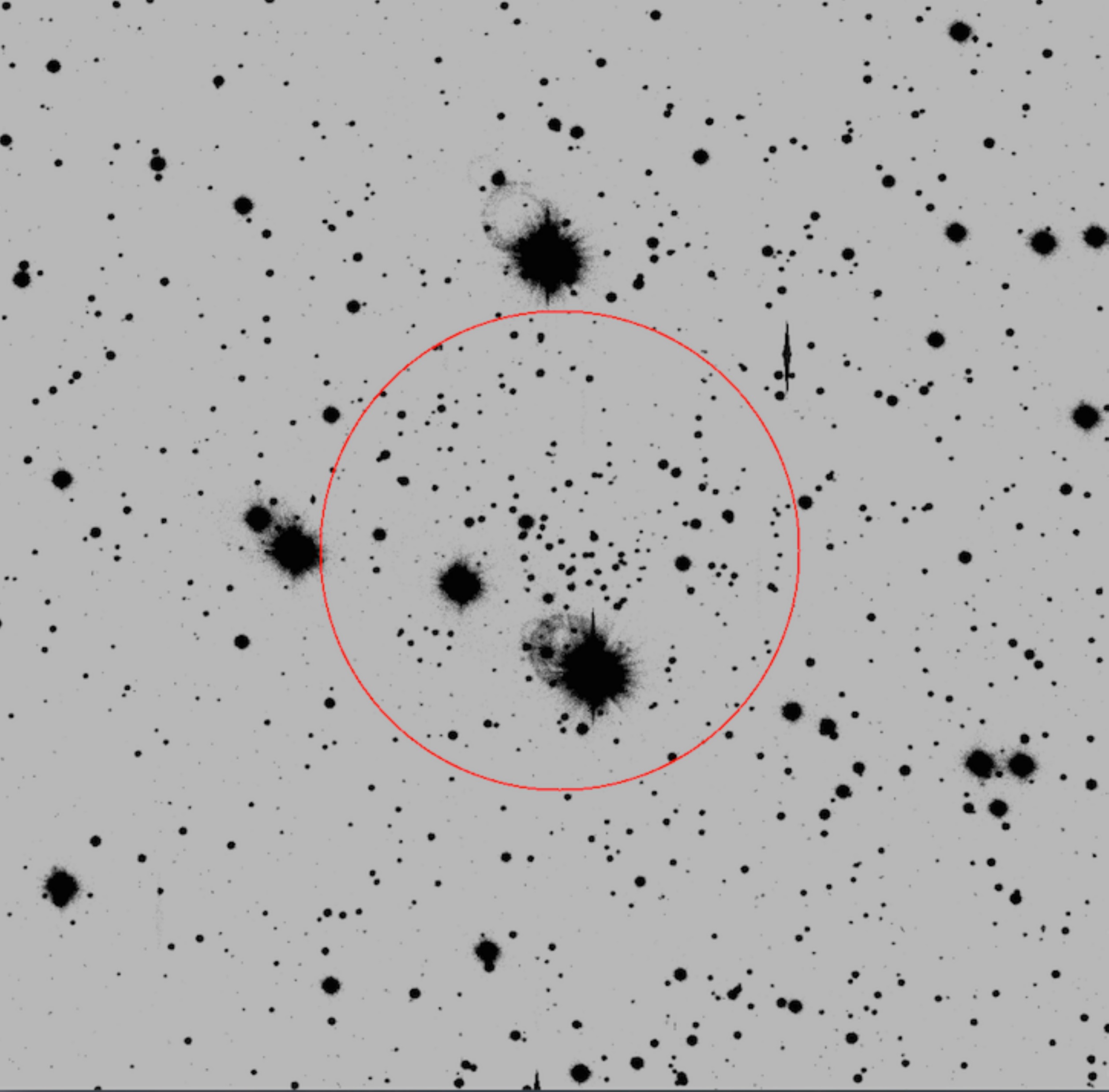}
\caption{$V$-band grey-scale images of the open cluster Ruprecht 6.
  North is at the top and east is to the left.
{\it Upper panel.} Y4KCam CCD image of the OC Ruprecht 6.
The field-of-view of the image is 
  $19.57\arcmin \times 19.57\arcmin$ ($4064 \times 4064$ pixels).
Thick red circle shows the region of Ruprecht 6 
  with radius of $2\arcmin$, centered on the coordinates of
  $\alpha_{J2000}$ = 06$^h$ 56$^m$ 06$^s$ and 
  $\delta_{J2000}$ = $-13$\arcdeg~ 15\arcmin~ 00\arcsec $ $
  from \citet{hase08}, which is adopted in this study.
Blue and white circles show the regions of radius of $2\arcmin$
  centered on the coordinates of \citet{dias02} 
  and SIMBAD, respectively.
{\it Lower panel.} Zoomed-in image of the region of Ruprecht 6.
Red circle shows the region of Ruprecht 6 with radius of 2\arcmin (the same as in the upper panel)
  and the field-of-view of the image is $\sim 9\arcmin \times 9\arcmin$.
\label{fig:vimage}}
\end{figure}

For the OC Ruprecht 6, we have obtained three 1200 sec, three 900 sec,
  and two 800 sec images for $B, V,$ and $I$ filters, respectively,
  under $1.5\arcsec$ median seeing conditions.
The final images are average combined for each filter.
Figure~\ref{fig:vimage} displays the combined grey-scale image of three $V$-band frames
  for the OC Ruprecht 6,
  where red circles show the region of Ruprecht 6 with radius of 2\arcmin.

The raw observation data have been processed with the 
  IRAF\footnote{IRAF is distributed by the National Optical Astronomy Observatory, 
  which is operated by the Association of Universities for Research in 
  Astronomy (AURA) under a cooperative agreement with 
  the National Science Foundation.}/\small{CCDRED} package
  with standard procedure. 
That is, the overscan correction, bias correction, and the twilight sky flattening
  have been made.
Since interference patterns are seen in the $I$-band images,
  we made $I$-band supersky image using all the $I$-band images obtained 
  on the same night and performed second flattening.
In addition, because of the large format of the detector and the slow speed of the shutter,
  the shutter shading correction is applied.
The IRAF script (y4kshut.cl) given at the CTIO homepage is used for the correction.

\begin{figure}
\centering
\includegraphics[width=90mm]{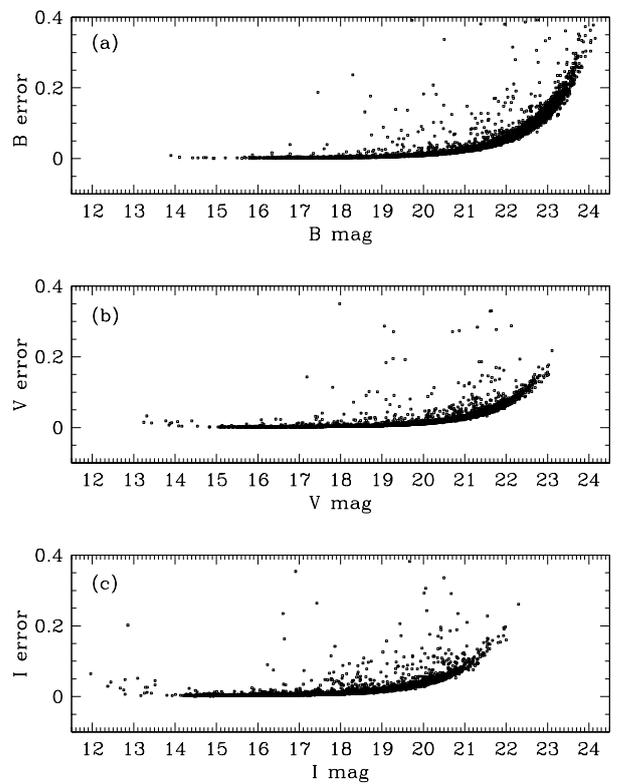}
\caption{Error distribution of the $B, V,$ and $I$-band photometry results
  as a function of magnitude. 
Only stars matched in all the three $B, V,$ and $I$-bands are plotted.
\label{fig:error}}
\end{figure}

Photometry was performed using the {\small DAOPHOT II}/{\small ALLSTAR} 
  stand-alone package 
  after separating the CCD quadrants to treat them as four separate CCD chips \citep{stetson90}.
This is also because each chip has different gain and readout noise.
Due to the large field of view,
  we adopt the quadratic variable point spread function.
Aperture corrections were obtained with $20-30$ isolated, bright, and unsaturated stars
 in each CCD quadrant.
Figure~\ref{fig:error} shows the error distribution 
  of the $BVI$ photometry results.
The photometric errors typically attain 0.1 mag at $B \approx 22.7$ mag,
  $V \approx 22.4$ mag and $I \approx 21.2$ mag \citep{kim09},
  indicating ten times of object signal above the background.

\begin{figure}
\centering
\includegraphics[width=80mm]{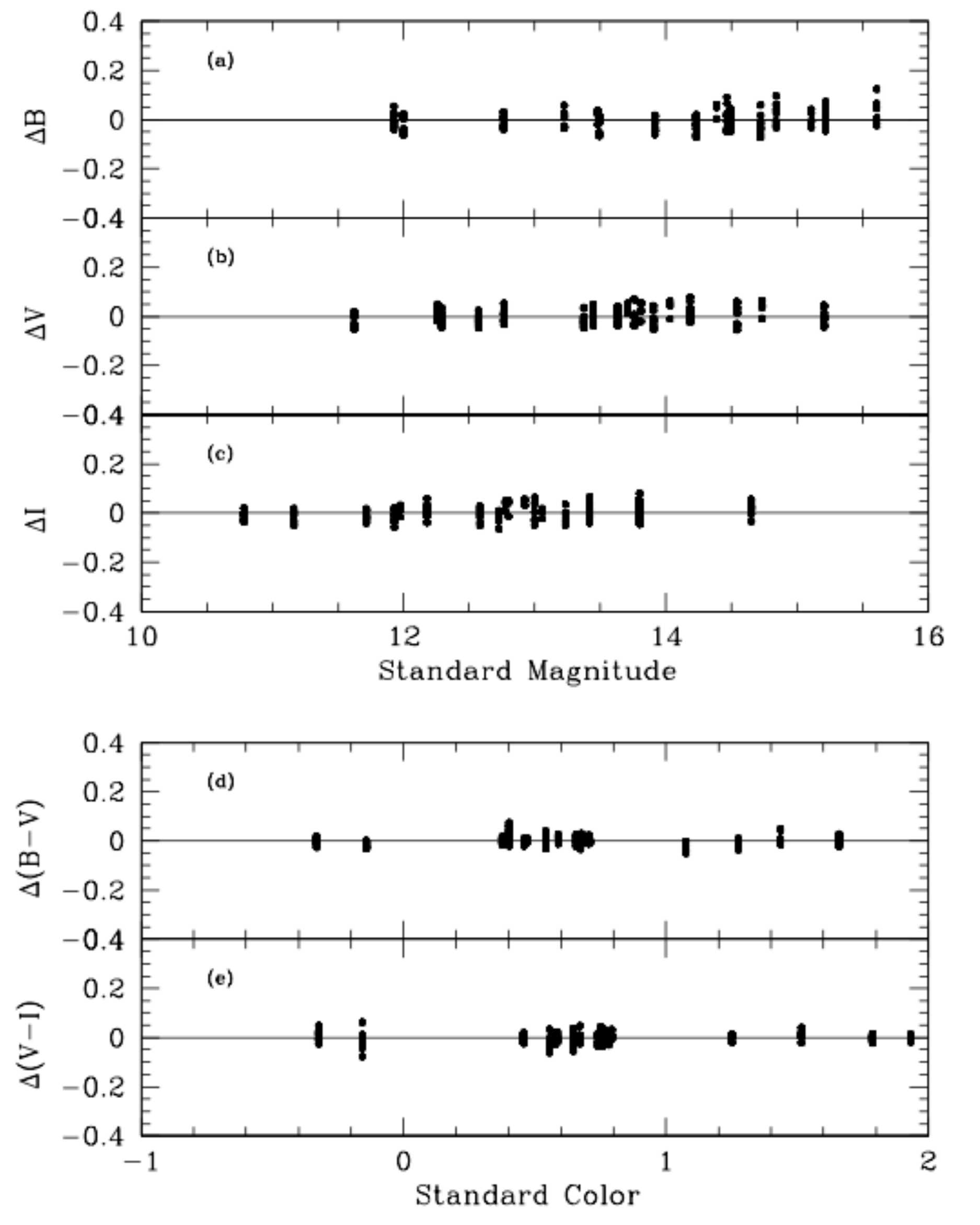}
\caption{$BVI$ residuals between standard and transformed magnitudes
of the standard stars, plotted against standard magnitudes. 
$\Delta$ means standard magnitude (color)
  minus transformed magnitude (color).
\label{fig:standardization}}
\end{figure}

Four \citet{landolt92,landolt07,landolt09} standard star fields 
  (PG0231+051, LB1735, LSS982, Rubin 149)
  were observed during the four nights of the observing run (2010 December $11-14$, UT)
  to obtain the standardization equations for this run and
  to convert the instrumental magnitudes to the standard magnitudes.
The transformation equations are

\begin{eqnarray*}
B = b - 0.285(\pm 0.009)~X_b - 0.127(\pm 0.005)(B-V) - 1.903(\pm 0.013) \\
V = v - 0.157(\pm 0.007)~X_v + 0.027(\pm 0.004)(B-V) - 1.693(\pm 0.011) \\
I = i - 0.056(\pm 0.007)~X_i + 0.019(\pm 0.003)(V-I) - 2.712(\pm 0.010) \\
\end{eqnarray*}
\noindent
  where small and capital letters represent instrumental and standard magnitudes, respectively,
  and $X$ is the airmass for each filter.
We also have obtained the secondary extinction coefficient ($k_{2B} = -0.052 \pm 0.016$)
  as a free parameter.
Including the secondary extinction coefficient, however, caused the errors
  of other parameters increase and the overall residual remained at the same
  value (0.037).
So we decided not to adopt $k_{2B}$ but follow the simple solution.
Figure~\ref{fig:standardization} (a), (b), and (c) show the standardization residuals
  for the magnitude differences, where
  the rms of the fits are $\Delta B=0.037$, $\Delta V=0.030$, $\Delta I=0.029$.
Figure~\ref{fig:standardization} (d) and (e) show the standardization residuals
  for the color differences, where
  the rms of the fits are $\Delta (B-V)=0.017$, and $\Delta (V-I)=0.019$.
Astrometry was done using the routines provided in astrometry.net
  \citep{lang10}.

The total number of stars with photometry is 5570,
while that for the region of Ruprecht 6 with radius $< 2\arcmin$ is 296.
Table~\ref{tab:phottab} lists the photometry of 92 stars of Ruprecht 6 
  with $V < 19$ mag.

\begin{table*}[t!]
\caption{Photometry data of the bright stars with $V<19$ mag in Ruprecht 6\label{tab:phottab}}
\centering
\setlength{\tabcolsep}{1.2mm}
\begin{tabular}{rcccccccccr}
\toprule
ID & R.A.(J2000) & Decl.(J2000)& $B$& $\sigma(B)$& $V$& $\sigma(V)$& $I$& $\sigma(I)$& $\chi$& 
  sharpness \\
    &  hh:mm:ss  & dd:mm:ss & & &   & &   & &   & \\
\midrule
  1& 06:55:58.41& $-$13:15:02.73& 20.25& 0.012& 18.99& 0.005& 17.46& 0.006& 1.010& $-$0.108 \\
  2& 06:55:58.57& $-$13:15:20.15& 19.97& 0.009& 18.95& 0.006& 17.68& 0.011& 1.312&  0.176 \\
  3& 06:55:59.86& $-$13:14:29.15& 20.07& 0.012& 18.97& 0.005& 17.59& 0.011& 1.062&  0.002 \\
  4& 06:55:59.96& $-$13:15:59.53& 20.11& 0.010& 18.80& 0.005& 17.37& 0.005& 1.004& $-$0.061 \\
  5& 06:55:59.97& $-$13:15:13.03& 19.00& 0.006& 17.91& 0.003& 16.62& 0.008& 1.032& $-$0.041 \\
  6& 06:56:00.08& $-$13:15:17.20& 20.15& 0.019& 18.75& 0.005& 17.18& 0.005& 1.180&  0.122 \\
  7& 06:56:00.13& $-$13:14:44.95& 19.22& 0.008& 18.21& 0.004& 16.97& 0.005& 0.704&  0.036 \\
  8& 06:56:00.20& $-$13:14:42.57& 16.64& 0.002& 15.71& 0.002& 14.62& 0.004& 1.335& $-$0.092 \\
  9& 06:56:00.32& $-$13:15:08.39& 19.65& 0.006& 18.75& 0.006& 17.56& 0.008& 1.198&  0.170 \\
 10& 06:56:00.51& $-$13:15:12.44& 19.95& 0.009& 18.60& 0.005& 17.11& 0.005& 1.084& $-$0.091 \\
 11& 06:56:00.61& $-$13:14:30.98& 19.34& 0.006& 18.50& 0.005& 17.40& 0.015& 1.152& $-$0.034 \\
 12& 06:56:00.98& $-$13:14:15.08& 19.36& 0.006& 18.51& 0.004& 17.40& 0.006& 1.071& $-$0.012 \\
 13& 06:56:01.08& $-$13:14:28.63& 18.47& 0.004& 17.59& 0.006& 16.40& 0.005& 1.184&  0.000 \\
 14& 06:56:01.19& $-$13:13:53.88& 19.29& 0.007& 18.45& 0.006& 17.39& 0.006& 1.242&  0.000 \\
 15& 06:56:01.21& $-$13:14:02.25& 18.27& 0.004& 17.37& 0.003& 16.17& 0.005& 1.275& $-$0.082 \\
 16& 06:56:01.47& $-$13:15:20.49& 18.71& 0.004& 17.88& 0.003& 16.84& 0.006& 1.181& $-$0.049 \\
 17& 06:56:01.65& $-$13:14:46.83& 19.85& 0.009& 18.96& 0.007& 17.79& 0.013& 1.495&  0.382 \\
 18& 06:56:02.10& $-$13:15:55.44& 18.96& 0.004& 18.15& 0.003& 17.09& 0.005& 1.008& $-$0.014 \\
 19& 06:56:02.11& $-$13:16:43.81& 17.84& 0.003& 17.02& 0.002& 16.04& 0.004& 1.110&  0.049 \\
 20& 06:56:02.26& $-$13:16:00.20& 19.88& 0.010& 18.60& 0.005& 17.19& 0.006& 1.120& $-$0.092 \\
 21& 06:56:02.41& $-$13:14:17.13& 17.09& 0.002& 16.17& 0.002& 15.14& 0.003& 1.198& $-$0.050 \\
 22& 06:56:03.23& $-$13:15:15.03& 20.38& 0.013& 18.89& 0.005& 17.17& 0.007& 1.200& $-$0.092 \\
 23& 06:56:03.85& $-$13:15:02.43& 19.38& 0.006& 18.51& 0.004& 17.41& 0.007& 1.073&  0.010 \\
 24& 06:56:03.99& $-$13:15:28.30& 18.79& 0.008& 17.95& 0.009& 16.77& 0.007& 1.288&  0.172 \\
 25& 06:56:04.02& $-$13:16:35.93& 19.78& 0.008& 18.70& 0.004& 17.44& 0.006& 1.026&  0.045 \\
 26& 06:56:04.04& $-$13:15:08.75& 18.86& 0.005& 17.99& 0.003& 16.83& 0.005& 1.175& $-$0.025 \\
 27& 06:56:04.34& $-$13:15:24.36& 20.15& 0.012& 18.70& 0.005& 16.89& 0.005& 1.165&  0.016 \\
 28& 06:56:04.46& $-$13:15:18.99& 19.32& 0.006& 18.25& 0.004& 17.00& 0.004& 1.022& $-$0.018 \\
 29& 06:56:04.67& $-$13:15:10.95& 19.59& 0.059& 18.70& 0.027& 17.48& 0.020& 1.059&  0.008 \\
 30& 06:56:04.77& $-$13:14:22.56& 19.60& 0.007& 18.36& 0.004& 16.95& 0.005& 1.069& $-$0.072 \\
 31& 06:56:04.79& $-$13:14:53.61& 18.05& 0.003& 17.21& 0.002& 16.10& 0.004& 1.152& $-$0.025 \\
 32& 06:56:04.86& $-$13:15:02.45& 19.55& 0.007& 18.68& 0.005& 17.55& 0.007& 1.071&  0.006 \\
 33& 06:56:04.94& $-$13:13:34.16& 18.45& 0.004& 17.29& 0.003& 15.98& 0.003& 1.111& $-$0.037 \\
 34& 06:56:04.96& $-$13:15:16.51& 19.15& 0.006& 18.20& 0.004& 17.01& 0.005& 1.147&  0.014 \\
 35& 06:56:05.07& $-$13:15:02.69& 19.58& 0.007& 18.71& 0.005& 17.56& 0.007& 1.082&  0.009 \\
 36& 06:56:05.21& $-$13:16:29.75& 16.74& 0.002& 16.20& 0.002& 15.52& 0.003& 1.084& $-$0.031 \\
 37& 06:56:05.51& $-$13:13:29.11& 19.69& 0.009& 18.49& 0.005& 17.11& 0.004& 1.089& $-$0.050 \\
 38& 06:56:05.53& $-$13:15:09.85& 19.45& 0.008& 18.60& 0.005& 17.47& 0.007& 0.921& $-$0.025 \\
 39& 06:56:05.57& $-$13:15:18.62& 18.47& 0.004& 17.62& 0.003& 16.52& 0.005& 1.142& $-$0.035 \\
 40& 06:56:05.65& $-$13:15:11.24& 19.32& 0.007& 18.46& 0.004& 17.33& 0.006& 1.103& $-$0.004 \\
 41& 06:56:05.66& $-$13:14:26.70& 18.30& 0.003& 16.89& 0.003& 15.26& 0.004& 1.273& $-$0.074 \\
 42& 06:56:05.80& $-$13:16:23.78& 18.84& 0.004& 18.02& 0.003& 16.94& 0.005& 1.095&  0.046 \\
 43& 06:56:05.85& $-$13:16:30.40& 19.67& 0.023& 18.75& 0.005& 17.54& 0.007& 1.769&  0.413 \\
 44& 06:56:05.91& $-$13:15:08.33& 19.10& 0.015& 18.21& 0.004& 17.05& 0.005& 1.701&  0.493 \\
 45& 06:56:05.91& $-$13:14:55.77& 19.52& 0.033& 18.63& 0.020& 17.48& 0.007& 2.310& $-$0.213 \\
 46& 06:56:06.16& $-$13:14:29.79& 19.81& 0.008& 18.91& 0.006& 17.75& 0.007& 1.073& $-$0.042 \\
 47& 06:56:06.17& $-$13:14:06.92& 19.54& 0.007& 18.40& 0.004& 17.05& 0.005& 1.087& $-$0.039 \\
 48& 06:56:06.32& $-$13:14:58.41& 18.88& 0.007& 18.05& 0.006& 16.97& 0.006& 1.180&  0.153 \\
 49& 06:56:06.37& $-$13:15:24.27& 17.25& 0.002& 15.86& 0.003& 14.24& 0.004& 1.603& $-$0.126 \\
 50& 06:56:06.41& $-$13:15:51.58& 16.56& 0.002& 15.14& 0.002& 13.52& 0.045& 6.578&  0.075 \\
 51& 06:56:06.51& $-$13:14:43.98& 18.95& 0.005& 18.03& 0.005& 16.83& 0.006& 1.353&  0.080 \\
 52& 06:56:06.61& $-$13:13:15.27& 20.35& 0.015& 18.88& 0.007& 17.20& 0.006& 1.025&  0.008 \\
 53& 06:56:06.62& $-$13:13:12.91& 18.71& 0.005& 17.82& 0.003& 16.65& 0.005& 1.157& $-$0.038 \\
 54& 06:56:06.63& $-$13:13:31.33& 17.93& 0.003& 16.92& 0.003& 15.63& 0.004& 1.279& $-$0.105 \\
 55& 06:56:06.67& $-$13:14:53.42& 19.84& 0.010& 18.56& 0.005& 17.13& 0.006& 1.227& $-$0.083 \\
 56& 06:56:06.95& $-$13:15:13.87& 19.15& 0.005& 18.28& 0.004& 17.08& 0.007& 1.035&  0.009 \\
 57& 06:56:07.03& $-$13:16:04.54& 19.68& 0.008& 18.82& 0.005& 17.67& 0.007& 1.083& $-$0.030 \\
 58& 06:56:07.12& $-$13:15:48.73& 18.75& 0.006& 17.88& 0.003& 16.74& 0.004& 1.192&  0.111 \\
 59& 06:56:07.21& $-$13:14:52.38& 18.99& 0.008& 18.08& 0.004& 17.03& 0.005& 0.918&  0.090 \\
 60& 06:56:07.28& $-$13:14:53.50& 19.53& 0.010& 18.57& 0.005& 17.32& 0.006& 0.909&  0.052 \\
 61& 06:56:07.46& $-$13:14:32.94& 19.60& 0.014& 18.41& 0.004& 16.97& 0.005& 1.430&  0.009 \\
 62& 06:56:07.53& $-$13:13:12.23& 18.18& 0.003& 17.29& 0.003& 16.12& 0.004& 1.126& $-$0.071 \\
 63& 06:56:07.57& $-$13:13:40.53& 19.14& 0.005& 18.32& 0.004& 17.17& 0.006& 1.155& $-$0.025 \\
 64& 06:56:07.58& $-$13:14:39.80& 18.88& 0.005& 18.03& 0.004& 16.90& 0.005& 1.103&  0.021 \\
 65& 06:56:07.63& $-$13:14:23.84& 19.31& 0.006& 18.46& 0.004& 17.32& 0.005& 1.028& $-$0.076 \\
\bottomrule
\end{tabular}
\end{table*}

\begin{table*}[t!]
\caption{Photometry data of the bright stars with $V<19$ mag in Ruprecht 6}
\centering
\setlength{\tabcolsep}{1.2mm}
\begin{tabular}{rcccccccccr}
\toprule
ID & R.A.(J2000) & Decl.(J2000)& $B$& $\sigma(B)$& $V$& $\sigma(V)$& $I$& $\sigma(I)$& $\chi$& 
  sharpness \\
    &  hh:mm:ss  & dd:mm:ss & & &   & &   & &   & \\
\midrule
 66& 06:56:07.93& $-$13:13:37.70& 19.65& 0.007& 18.76& 0.005& 17.60& 0.006& 1.025& $-$0.065 \\
 67& 06:56:08.14& $-$13:16:27.92& 19.95& 0.010& 18.60& 0.005& 16.97& 0.005& 1.029& $-$0.058 \\
 68& 06:56:08.23& $-$13:14:37.70& 19.91& 0.010& 18.75& 0.005& 17.39& 0.006& 1.089&  0.004 \\
 69& 06:56:08.44& $-$13:16:27.10& 18.09& 0.003& 17.21& 0.003& 16.03& 0.004& 1.099& $-$0.030 \\
 70& 06:56:08.65& $-$13:14:44.54& 18.85& 0.005& 17.97& 0.003& 16.84& 0.004& 1.001& $-$0.021 \\
 71& 06:56:09.07& $-$13:14:46.01& 17.22& 0.002& 16.37& 0.003& 15.30& 0.003& 1.372& $-$0.033 \\
 72& 06:56:09.56& $-$13:13:57.52& 19.16& 0.005& 18.51& 0.005& 17.65& 0.008& 1.169& $-$0.114 \\
 73& 06:56:09.62& $-$13:15:22.41& 19.26& 0.047& 18.36& 0.021& 17.19& 0.012& 4.276&  2.010 \\
 74& 06:56:09.64& $-$13:16:32.90& 17.36& 0.002& 16.07& 0.002& 14.55& 0.003& 1.227& $-$0.073 \\
 75& 06:56:09.71& $-$13:15:42.04& 19.68& 0.008& 18.81& 0.005& 17.68& 0.006& 1.058&  0.098 \\
 76& 06:56:09.72& $-$13:14:09.50& 19.82& 0.008& 18.76& 0.005& 17.44& 0.006& 1.042&  0.066 \\
 77& 06:56:10.02& $-$13:13:49.08& 18.05& 0.003& 17.10& 0.003& 15.93& 0.003& 1.137& $-$0.077 \\
 78& 06:56:10.05& $-$13:13:42.90& 19.45& 0.006& 18.58& 0.004& 17.45& 0.005& 1.004& $-$0.012 \\
 79& 06:56:10.24& $-$13:15:41.29& 19.31& 0.007& 18.43& 0.004& 17.31& 0.007& 1.135& $-$0.014 \\
 80& 06:56:10.24& $-$13:14:28.93& 19.42& 0.008& 18.55& 0.006& 17.38& 0.007& 1.100&  0.082 \\
 81& 06:56:10.50& $-$13:13:56.56& 20.10& 0.010& 18.63& 0.005& 16.88& 0.004& 1.047& $-$0.175 \\
 82& 06:56:10.76& $-$13:15:55.35& 20.06& 0.010& 18.73& 0.005& 17.26& 0.006& 1.127& $-$0.034 \\
 83& 06:56:11.08& $-$13:15:46.69& 19.38& 0.006& 18.52& 0.004& 17.45& 0.005& 0.985&  0.024 \\
 84& 06:56:11.30& $-$13:14:25.49& 17.47& 0.003& 16.08& 0.002& 14.44& 0.003& 1.214& $-$0.058 \\
 85& 06:56:11.40& $-$13:13:52.18& 17.95& 0.003& 17.08& 0.002& 15.96& 0.004& 1.198&  0.011 \\
 86& 06:56:11.41& $-$13:15:40.76& 19.71& 0.007& 18.82& 0.005& 17.68& 0.008& 1.092& $-$0.053 \\
 87& 06:56:11.44& $-$13:14:24.73& 19.82& 0.012& 18.90& 0.009& 17.66& 0.012& 0.598& $-$0.023 \\
 88& 06:56:11.96& $-$13:14:12.16& 17.24& 0.002& 16.35& 0.002& 15.20& 0.003& 1.199& $-$0.021 \\
 89& 06:56:12.02& $-$13:13:41.90& 19.22& 0.006& 18.36& 0.004& 17.21& 0.005& 1.047& $-$0.025 \\
 90& 06:56:12.16& $-$13:14:52.31& 16.12& 0.002& 15.40& 0.002& 14.54& 0.003& 1.330& $-$0.083 \\
 91& 06:56:12.16& $-$13:14:14.52& 19.92& 0.010& 18.80& 0.005& 17.44& 0.006& 0.954& $-$0.010 \\
 92& 06:56:12.26& $-$13:15:10.06& 18.89& 0.005& 17.72& 0.003& 16.38& 0.004& 1.174& $-$0.086 \\
\bottomrule
\end{tabular}
\small{ \\
}
\end{table*}

\section{Results\label{sec:results}}
  \subsection{Cluster Center and Size\label{sec:center}}

For the center coordinates of the OC Ruprecht 6, 
  \citet{hase08} presents $\alpha_{J2000}$ = 06$^h$ 56$^m$ 06$^s$, 
  $\delta_{J2000}$ = $-13$\arcdeg~ 15\arcmin~ 00\arcsec,
  which we adopt in this study.
While 
  \citet{dias02} (Version 3.5, 2016 January 28) lists 
  $\alpha_{J2000}$ = $06^h$ 56$^m$ 00$^s$, 
  $\delta_{J2000}$ = $-13\arcdeg~ 17\arcmin~ 00\arcsec$ and
  SIMBAD gives $\alpha_{J2000}$ = 06$^h$ 55$^m$ 55.2$^s$, 
  $\delta_{J2000}$ = $-13$\arcdeg~ 16\arcmin~ 48\arcsec,
  both of these values result in poor identifications of the cluster center.
The coordinates given by \citet{hase08} 
  point to the cluster center more accurately as seen in 
  Figure~\ref{fig:vimage}.

\begin{figure}
\centering
\includegraphics[width=80mm]{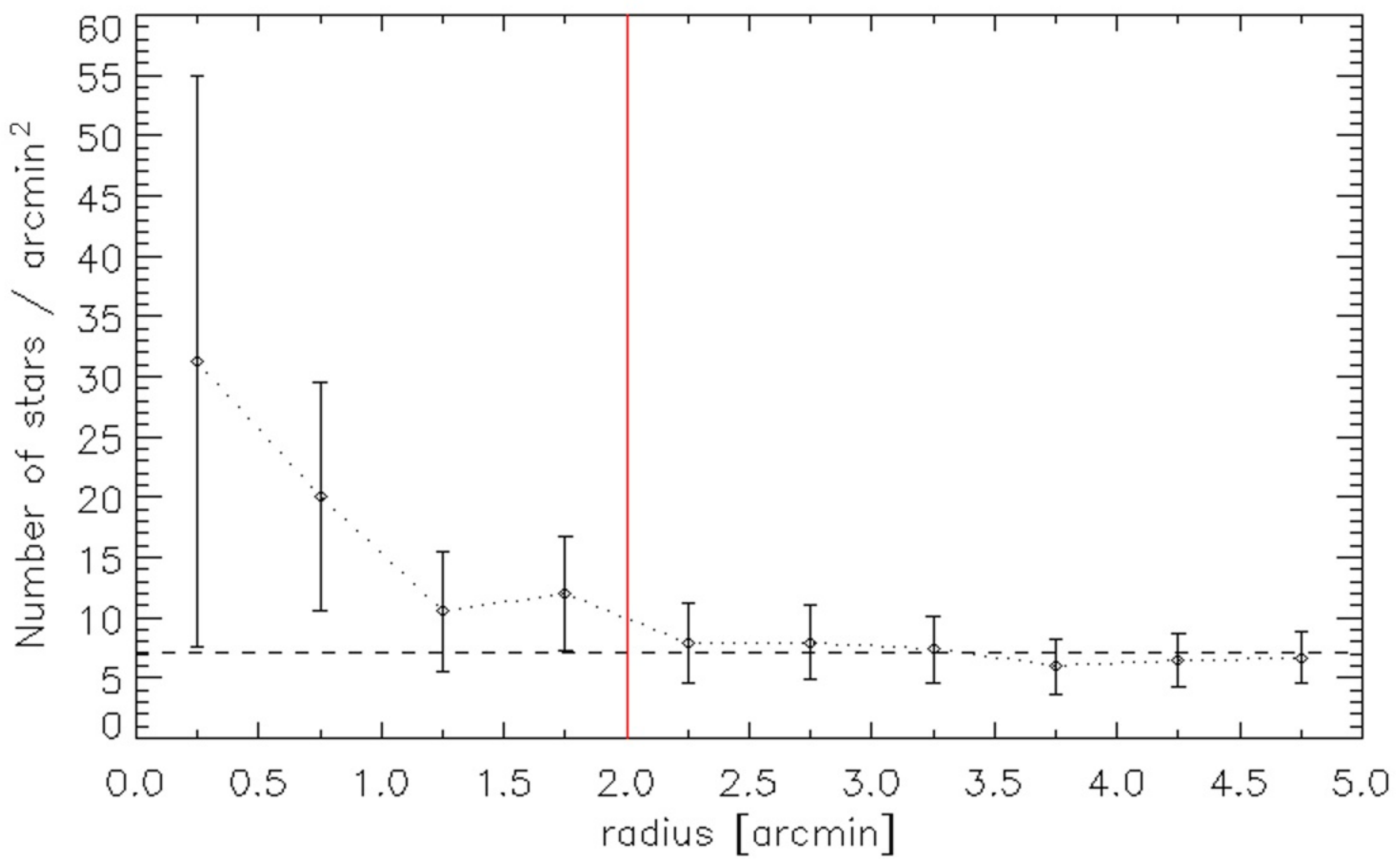}
\caption{Radial number density profile for the stars with $V \le 20$ mag in the field
  of the OC Ruprecht 6 with bin size of $0.5\arcmin$.
The vertical red solid line represents our adopted size (radius) of the cluster 
  ($2\arcmin$).
Error bars denote the Poisson errors and the horizontal dashed line
  the level of field stellar density.
\label{fig:size}}
\end{figure}

Using the bright stars ($V \le 20$ mag) with the cluster center fixed as above,
  we plotted the radial number density profile with the bin size of $0.5\arcmin$
  in Figure~\ref{fig:size}.
We determined the radius of Ruprecht 6 to be $2.0\arcmin$,
  in which radius most of the member stars are located.

  \subsection{Color-Magnitude Diagrams\label{sec:CMDs}}
\begin{figure*}
\centering
\includegraphics[width=160mm]{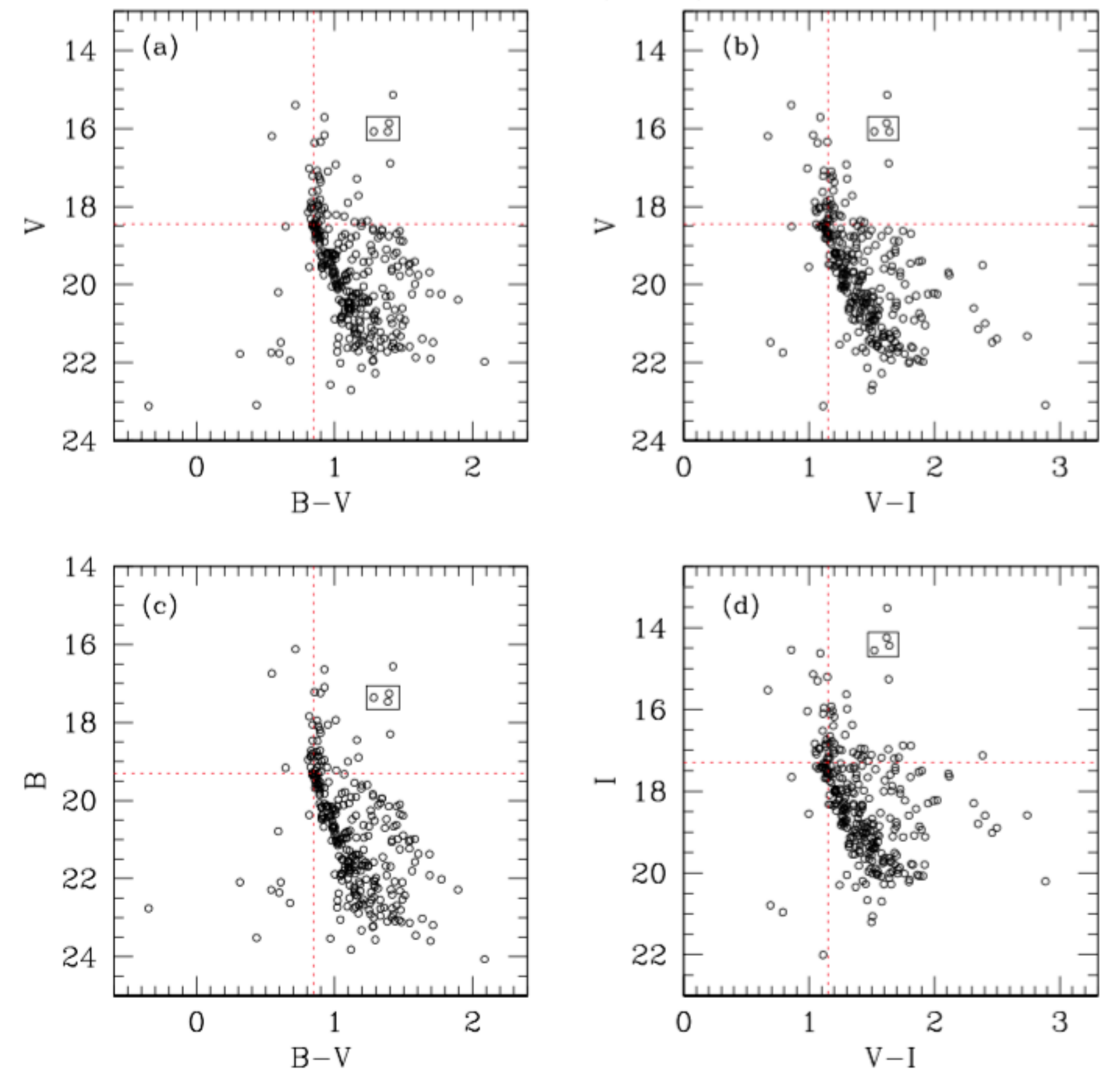}
\caption{(a) $V-(B-V)$, (b) $V-(V-I)$, (c) $B-(B-V)$, and (d) $I-(V-I)$ color-magnitude diagrams
for the 296 stars of the OC Ruprecht 6 at radius $<2\arcmin$.
Red dotted lines show the turn-off point at $B\approx 19.30$, $V\approx 18.45$, and
  $I\approx 17.30$ mag.
Boxes show the location of the three red clump (RC) stars.
Electronic file for the photometry of all the 296 stars is provided via JKAS.
\label{fig:cmd1}}
\end{figure*}

\begin{figure*}
\centering
\includegraphics[width=160mm]{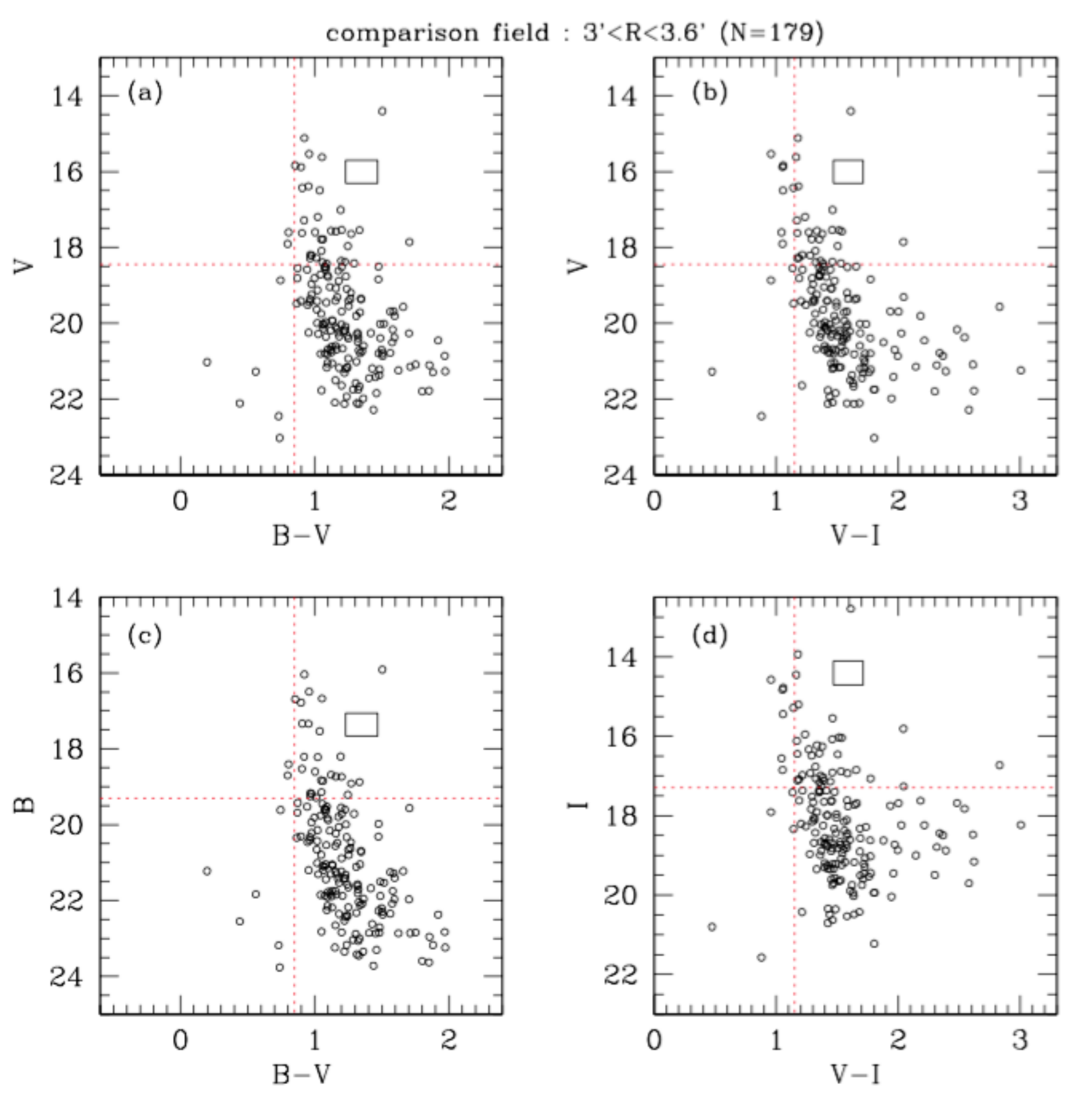}
\caption{Same as in Figure~\ref{fig:cmd1}, but for the comparison field
  of $3\arcmin <$ radius $<3.6\arcmin$ around Ruprecht 6.
This radius range gives the same area as in the case of Ruprecht 6.
The red dotted lines and black boxes are the turn-off point and 
  the region of the RC stars for Ruprecht 6
  shown here just for guidelines.
\label{fig:cmd1comp}}
\end{figure*}

Figure~\ref{fig:cmd1} shows the $V-(B-V)$ (panel a), $V-(V-I)$ (panel b), $B-(B-V)$ (panel c), 
  and $I-(V-I)$ (panel d) color-magnitude diagrams (CMDs) of the OC Ruprecht 6 
  for the radial range of $< 2\arcmin$ obtained in this study,
  while Figure~\ref{fig:cmd1comp} shows the comparison region ($3\arcmin < R
  < 3.6\arcmin$) with the same area.
Some characteristic features that can be found for Ruprecht 6 in these CMDs are :
(i) main-sequence stars are clearly seen 
  and red dotted lines show the blue turn-off point 
  at $B\approx 19.30$ mag, $V\approx18.45$ mag,
  $I\approx17.30$ mag, $B-V \approx 0.85$ mag, and $V-I \approx 1.15$ mag
  (see Figure 2 of 
  Kaluzny (1994)
  for the definition of the blue turn-off),
(ii) some red giant stars evolved after the turn-off are found, and
(iii) three RC stars are seen and denoted as boxes in Figure~\ref{fig:cmd1}
  with mean magnitudes and colors of 
  $B=17.36 \pm 0.09$ mag, $V=16.00 \pm 0.10$ mag, $I=14.41 \pm 0.13$ mag,
  $B-V=1.35 \pm0.05$, mag $V-I=1.59 \pm0.05$ mag.

\begin{figure}
\centering
\includegraphics[width=80mm]{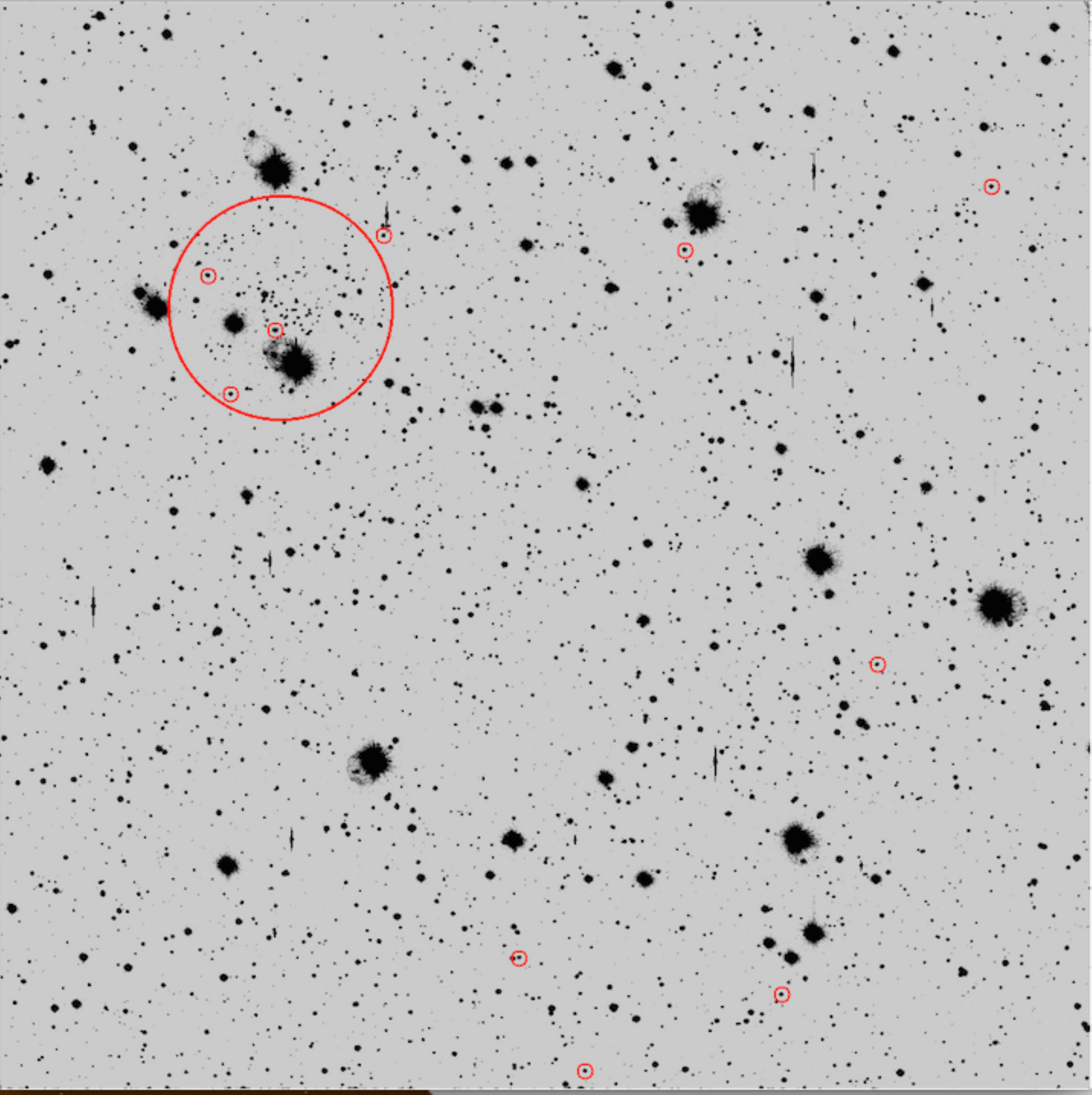}
\caption{$V$-band grey-scale images of the open cluster Ruprecht 6
  with the thick and large, red circle 
  same as that in Figure~\ref{fig:vimage} (radius of $2\arcmin$).
Small red circles show the 10 stars with magnitudes and colors
  located in the box area of the RC stars in the CMDs of
  Figure~\ref{fig:cmd1} and Figure~\ref{fig:cmd1comp}.
Note that three stars, out of 10, are located in the 
  small area of the radial content of Ruprecht 6.
\label{fig:starsRCarea}}
\end{figure}

To further check the membership probability of the three RC stars statistically,
  we extracted all stars with similar magnitudes and colors to the RC stars
  in the whole observed area of $19.57\arcmin \times 19.57\arcmin$.
Figure~\ref{fig:starsRCarea} shows the spatial distribution of
  the resultant 10 stars located in the RC box area of 
  Figure~\ref{fig:cmd1} and Figure~\ref{fig:cmd1comp},
  including the three RC stars in Ruprecht 6.
The stellar spatial density of the three stars in the cluster area is
  $3 / (\pi 2^2) = 0.239 ~{\rm arcmin}^{-2}$,
while that of the seven stars outside of the cluster area
  is $7 / (19.57^2 - \pi 2^2) = 0.019 ~{\rm arcmin}^{-2}$.
The stellar density of the cluster area is
  $0.239 / 0.019 = 12.58$ times more higher than that of the background area,
  and thus we conclude that the three RC stars in Ruprecht 6 are
  members of the cluster with high probability.

  \subsection{Distance Estimation\label{sec:distance}}
RC giant stars are low-mass stars at the evolutionary stage of core-helium-burning,
  that define a sharpest feature (almost constant absolute magnitude) 
  in the CMDs of stellar systems like nearby galaxies and star clusters
  (\citet{pac98,kim03,kim05,kyeong11,kara13,girardi16,davies17}, but,
  see also \citet{wan15}).
In the near-infrared (NIR), especially in the $K_S$-band, the RC is known to 
  have small dependence on metallicity and age, showing
  $M_{K_S} = -1.595 \pm 0.025$ mag and an intrinsic color of 
  $(J-K_S)_0 = 0.612 \pm 0.003$ mag \citep{Yaze13,ozd16}.
Recently, \citet{girardi16} published a detailed review paper on the RC stars,
  and summarized the mean absolute magnitudes of these stars,
  which could possibly be used to estimate the distances and extinctions
  to stellar systems with the age of 1 to 10 Gyr
  though with some caveats such as population effects.
Table~\ref{tab:RCmag} shows the mean $I$-band absolute magnitudes of RC stars
  from \citet{girardi16} and 
  the mean values of the five data is $\langle M_I \rangle = -0.236 \pm 0.024$ mag.

Since it is known that the near-infrared $K$-band magnitude of the RC stars is
  not sensitive to the age and metallicity of the star cluster
  than the optical $I$-band magnitude \citep{saraje99,grochol02,kyeong11},
  we mainly used the mean $K$(RC) magnitude of the RC stars in Ruprecht 6.
Table~\ref{tab:RCJK} shows the $JHK_s$ magnitudes of the three RC stars in Ruprecht 6
  extracted from the 2MASS (Two Micron All Sky Survey\footnote{available at
  http://www.ipac.caltech.edu/2mass/releases/allsky/},
  \citet{skrut97,skrut06}) data archive.
Using the mean $K$(RC) magnitude of $\langle K_s \rangle = 12.39 \pm 0.21$,
  the extinction value of $A_V = 1.30$ (see Section~\ref{sec:reddening}), and
  the extinction ratio of $A_K = 0.118 A_V = 0.15$ \citep{dutra02},
  we obtain $(m-M)_0 = K_s - M_{K_s} - A_K = 12.39 - (-1.595) - 0.15 = 
  13.84 \pm 0.21$ ($d=5.86 \pm 0.60$).

\begin{table*}[t!]
\caption{Mean $I$-band absolute magnitudes of RC stars$^{\dagger}$\label{tab:RCmag}}
\centering
\setlength{\tabcolsep}{1.2mm}
\begin{tabular}{lll}
\toprule
$M_I$ & Reference & Comments\\
\midrule
$-0.279 \pm 0.088$ & \citet{pac98} & Includes reddening correction \\
$-0.23  \pm 0.03 $ & \citet{sta98} & Volume limited to 70 pc \\
$-0.209          $ & \citet{gir98} & Lutz-Kelker bias corrected \\
$-0.22  \pm 0.03 $ & \citet{gro08} & First uses revised {\it Hipparcos} parallaxes \\
$-0.24  \pm 0.01 $ & \citet{fran14} & Peak of luminosity distribution \\
\bottomrule
\end{tabular}
\small{\\
Note.-- $^{\dagger}$From \citet{girardi16}. Values generally from {\it Hipparcos}. }
\end{table*}

\begin{table*}[t!]
\caption{2MASS Photometry of the RC stars in Ruprecht 6\label{tab:RCJK}}
\centering
\setlength{\tabcolsep}{1.2mm}
\begin{tabular}{cccccccccc}
\toprule
IDs$^{\dagger}$& R.A.(J2000) & Decl.(J2000)& $J$& $\sigma(J)$& $H$& $\sigma(H)$& $K_s$& $\sigma(K_s)$& $J-K_s$ \\
    &  hh:mm:ss  & dd:mm:ss & & &   & &   & &    \\
\midrule
 49 & 06:56:06.37& $-$13:15:24.27& 13.004& 0.023& 12.350& 0.026& 12.120& 0.024& 0.884\\
 74 & 06:56:09.64& $-$13:16:32.90& 13.383& 0.024& 12.764& 0.029& 12.634& 0.028& 0.784\\
 84 & 06:56:11.30& $-$13:14:25.49& 13.184& 0.026& 12.564& 0.027& 12.400& 0.029& 0.749\\
\bottomrule
\end{tabular}
\small{\\
Note.-- $^{\dagger}$IDs from Table~\ref{tab:phottab}.
}
\end{table*}

  \subsection{Reddening Estimation\label{sec:reddening}}
For the estimation of the interstellar reddening value toward the OC Ruprecht 6,
  we adopted the following four methods.
First, we have used the mean NIR color of the RC stars
  ($\langle J-K_s \rangle = 0.81 \pm 0.06$) from Table~\ref{tab:RCJK} and
  the intrinsic color of RC stars ($(J-K_s)_0 = 0.612 \pm 0.003$ mag)
  mentioned in Section~\ref{sec:distance},
  we obtain $E(J-K_s) = (J-K_s) - (J-K_s)_0 = 0.20 \pm 0.06$.
Using the relation $E(J-K_s) = 0.488 E(B-V)$ \citep{kim06},
  we obtain $E(B-V) = 0.41 \pm 0.06$.

Secondly, we have used the mean optical color of the RC stars in Ruprecht 6: 
  $B-V = 1.35 \pm 0.05$ (Section~\ref{sec:CMDs}).
  \citet{janes94} gave the mean color and magnitude of the RC stars 
  in old OCs as $(B-V)_{0,RC} = 0.95 \pm 0.10$ and $M_{V,RC} = 0.90 \pm 0.40$,
  when the $V$ magnitude difference between the RC stars and 
  the main sequence turn-off of the clusters, $\delta V$, is greater than one
  \citep{kim03}.
Since $\delta V \approx 18.45 - 16.00 > 1 $ for Ruprecht 6,
  we obtain $E(B-V) = (B-V) - (B-V)_0 = 1.35 - 0.95 = 0.40 \pm 0.11$.

\begin{figure}
\centering
\includegraphics[width=80mm]{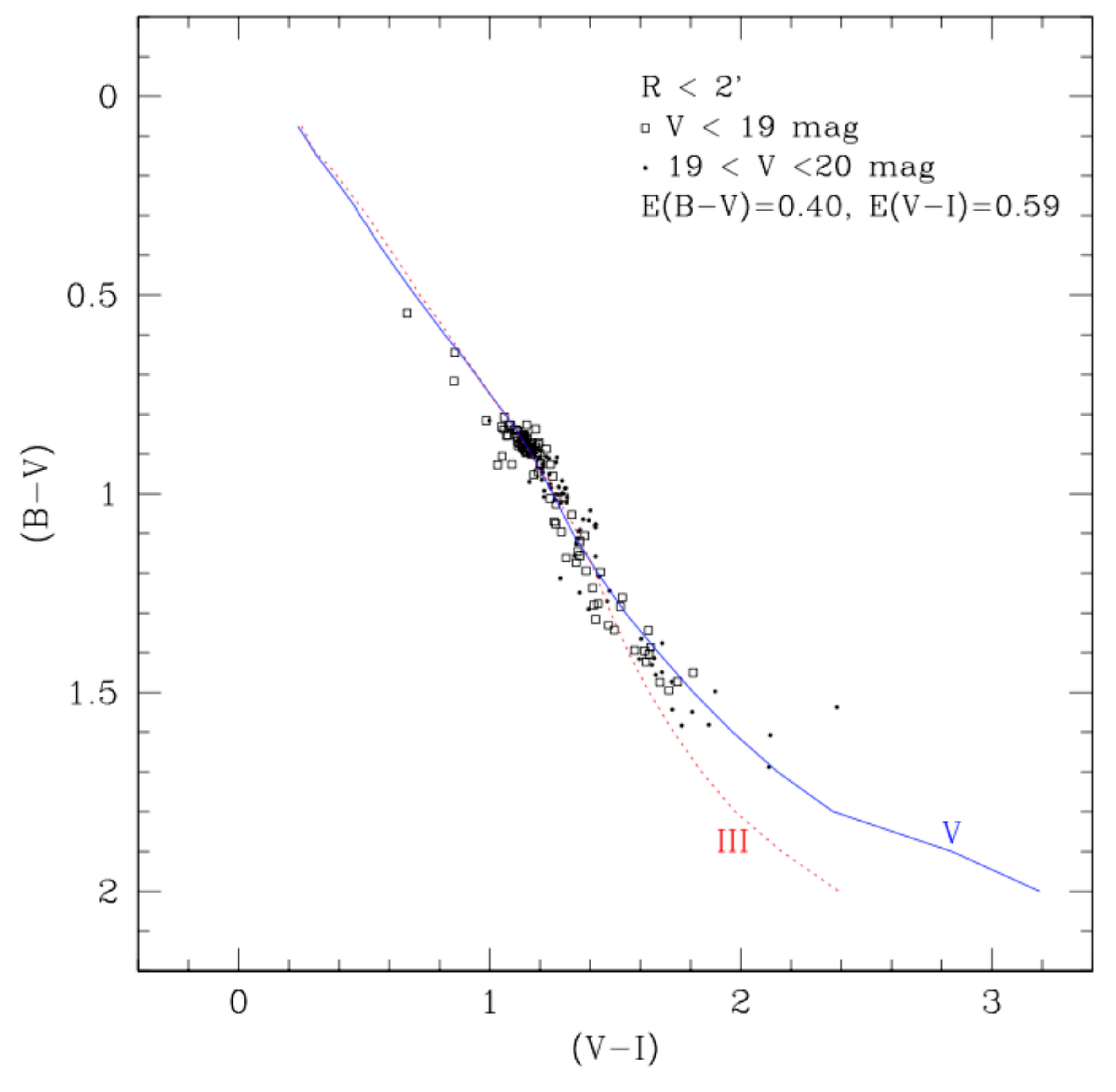}
\caption{$(B-V)$ versus $(V-I)$ diagram for the bright ($V < 19$ mag, open squares) and
  less bright ($19 < V < 20$ mag, dots) stars of the OC Ruprecht 6 at radius $<2\arcmin$.
Solid and dotted lines represent the intrinsic relations for the dwarf and giant stars, respectively, 
  from \citet{sung13}, which are shifted according to the reddenings of $E(B-V)=0.40$ and
  $E(V-I)=0.59$.
\label{fig:tcd}}
\end{figure}

Thirdly, we have used the $(B-V)$ versus $(V-I)$ diagram of 
  bright stars with $V < 19$ mag and
  stars of $19 < V < 20$ mag in Ruprecht 6 at radius $<2\arcmin$
  as shown in Figure~\ref{fig:tcd}.
Solid and dotted lines show the intrinsic relations 
  for the dwarf and giant stars, respectively,
  from \citet{sung13}, which are shifted according to the reddenings of $E(B-V)=0.40$ and
  $E(V-I)=0.59$.
Although the reddening vector in the diagram is very similar to the distribution
  of main-sequence and giant stars,
  we can get a hint of the amount of the reddening values from this diagram.
From the comparison of
  the observed colors of the bright stars of Ruprecht 6 and the intrinsic colors,
  we derive the reddening values of 
  $E(B-V)=0.40 \pm 0.10$ and 
  $E(V-I)=0.59 \pm 0.10$.

Finally, PARSEC isochrone fittings are made as shown in Section~\ref{sec:padova},
  which gives the best match with $E(B-V)=0.42$ and $E(V-I)=0.60$,
  which are in agreement with the values above within uncertainties.

Considering the results of these four methods,
  we assume that the true reddening values toward Ruprecht 6 are
  in the range of $E(B-V)=0.40 - 0.42$ and $E(V-I)=0.59 - 0.60$,
  and in this study we adopt the values of 
  $E(B-V)=0.42$ and $E(V-I)=0.60$ 
  which afford the best isochrone matches (see Section~\ref{sec:padova}).
This is in very good agreement with the value ($E(B-V)=0.43$)
  listed in the DAML02 catalog.
The interstellar extinction laws given by \citet{cardel89} are used to calculate the extinctions
  for other colors for the total-to-selective extinction ratio of $R_V=3.1$ : 
  $A_B= 4.14 E(B-V)= 1.74$,
  $A_V= 3.10 E(B-V)= 1.30$,
  $A_I= 1.48 E(B-V)= 0.62$
  \citep{leekim00, kim12}.

  \subsection{PARSEC Isochrone Fittings\label{sec:padova}}
\begin{figure*}
\centering
\includegraphics[width=160mm]{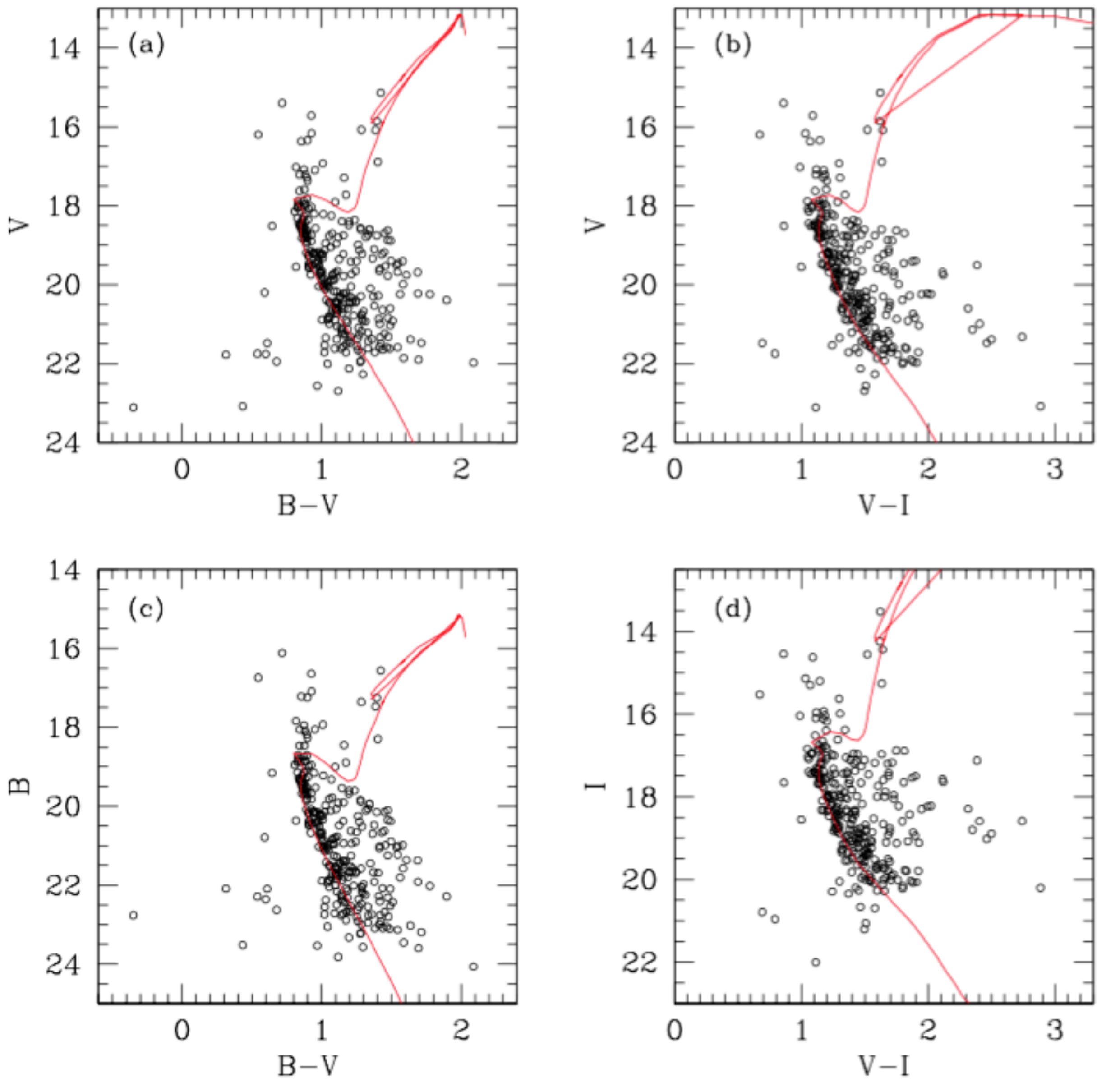}
\caption{PARSEC isochrone fittings onto the color-magnitude diagrams of Ruprecht 6
for (a) $V-(B-V)$, (b) $V-(V-I)$, (c) $B-(B-V)$, and (d) $I-(V-I)$ parameter space.
Open circles are the 296 stars of Ruprecht 6 at radius $<2\arcmin$.
The red solid lines show the PARSEC isochrones of log (age) $=9.50$ (t $=3.16$ Gyr)
  and [Fe/H] $= -0.42$ dex shifted according to $E(B-V) = 0.42$, $E(V-I) = 0.60$,
  $B - M_B = 15.58$, $V - M_V = 15.14$, and $I - M_I = 14.46$.
}
\label{fig:cmd_pdv}
\end{figure*}

In Figure~\ref{fig:cmd_pdv}, we have plotted the $V$ versus $(B-V)$ (a),
  $V$ versus $(V-I)$ (b), $B$ versus $(B-V)$ (c), and $I$ versus $(V-I)$ (d)
  CMDs for Ruprecht 6 together with the theoretical 
  PAdova and TRieste Stellar Evolution Code (PARSEC) isochrones 
 \citep{bertelli94,gir00,gir02,bre12}.
The best matched isochrones are for the parameters of
  log (age) $=9.50 \pm 0.10$ (t $=3.16 \pm 0.82$ Gyr)
  and [Fe/H] $= -0.42 \pm 0.04$ dex shifted according to $E(B-V) = 0.42$, $E(V-I) = 0.60$,
  $B - M_B = 15.58$, $V - M_V = 15.14$, and $I - M_I = 14.46$.
The metallicity and age values obtained in this study are 
  in excellent agreement with those of \citet{hase08} ([Fe/H] $= -0.41$ and log (t) $=9.50$, t $= 3.2$ Gyr)
  and DAML02 ([Fe/H] $= -0.38$ and log (t) $=9.50$).

With the limited data of $BVI$ photometry only,
  it is not an easy task to determine the reddening, distance, age, and
  metallicity simultaneously from the isochrone fitting
  due to the degeneracy of parameters.
We, therefore, have used pre-determined values of distance (Section~\ref{sec:distance})
  and reddening (Section~\ref{sec:reddening}) for the isochrone fittings
  just to find optimum values of age and metallicity,
  after which slight fine-tunings are made.
Good matches of the theoretical isochrones and the observed photometry data
  lend support for the distance and reddening values,
  and the newly obtained values of age and metallicity.

\section{Discussion\label{sec:discuss}}
  \subsection{Red Clump Stars\label{sec:RCstars}}

Among the 10 RC stars shown in Figure~\ref{fig:starsRCarea},
  one star is located just outside of the $2\arcmin$ radius circle of Ruprecht 6,
  with its coordinates R.A. $(J2000)=06^h$ 56$^m$ 6.366$^s$ and 
  Decl. $(J2000)=-13\arcdeg~ 15\arcmin~ 24.27\arcsec$.
Since it is very likely that this star could also be a member of the cluster
  considering its close location to the cluster area,
  we can include it in our estimation of the cluster parameters.
The optical and 2MASS NIR photometry results of the star are :
  $B = 17.358 \pm 0.003$ mag,
  $V = 15.912 \pm 0.002$ mag,
  $I = 14.273 \pm 0.004$ mag,
  $J = 12.981 \pm 0.026$ mag,
  $H = 12.229 \pm 0.022$ mag,
  $K_S = 12.065 \pm 0.024$ mag, and $(J-K_s) = 0.916$ mag.

Inclusion of this star results in the mean $K(RC)$ magnitude to be
  $\langle K_s \rangle = 12.31 \pm 0.23$ mag and
  this gives the distance modulus to be 
  $(m-M)_0 = K_s - M_{K_s} - A_K = 12.31 - (-1.595) - 0.15 = 
  13.76 \pm 0.23$ ($d=5.65 \pm 0.63$ kpc,
  still using the reddening values obtained in Section~\ref{sec:reddening}),
  which agrees within the error range
  but somewhat shorter than that in Section~\ref{sec:distance}.
The mean NIR color of the four RC stars including the above one is 
  $\langle (J-K_s) \rangle = 0.83 \pm 0.07$ mag and 
  this results in a little bit larger reddening values of
  $E(J-K_s) = (J-K_s) - (J-K_s)_0 = 0.22 \pm 0.07$ and 
  $E(B-V) = 0.45 \pm 0.07$.

  \subsection{Galactocentric Metallicity Distribution\label{sec:MetDist}}

OCs have been used as one of the tools to probe the Galactocentric radial metallicity
  distribution in our own Galaxy \citep{kim03,wu09,ryu11}.
This radial variation of the metallicity in the disk of the Galaxy is a powerful tool
  for the understanding of the star formation and chemical evolution of the system
  \citep{fern17}.
\citet{kim05} have compiled the slope ($=\Delta$[Fe/H]$/ \Delta R_{\mathrm{GC}}$)
  values of the Galactocentric radial metallicity gradient published up to then.
Using the nine published slope values, they obtained the mean value of the slope
  $\Delta$[Fe/H]$/ \Delta R_{\mathrm{GC}} = -0.066\pm0.019$.
In Table~\ref{tab:slope}, we have compiled again the slopes and intercept values
  incorporating recently published results for OCs.

\begin{figure}
\centering
\includegraphics[width=80mm]{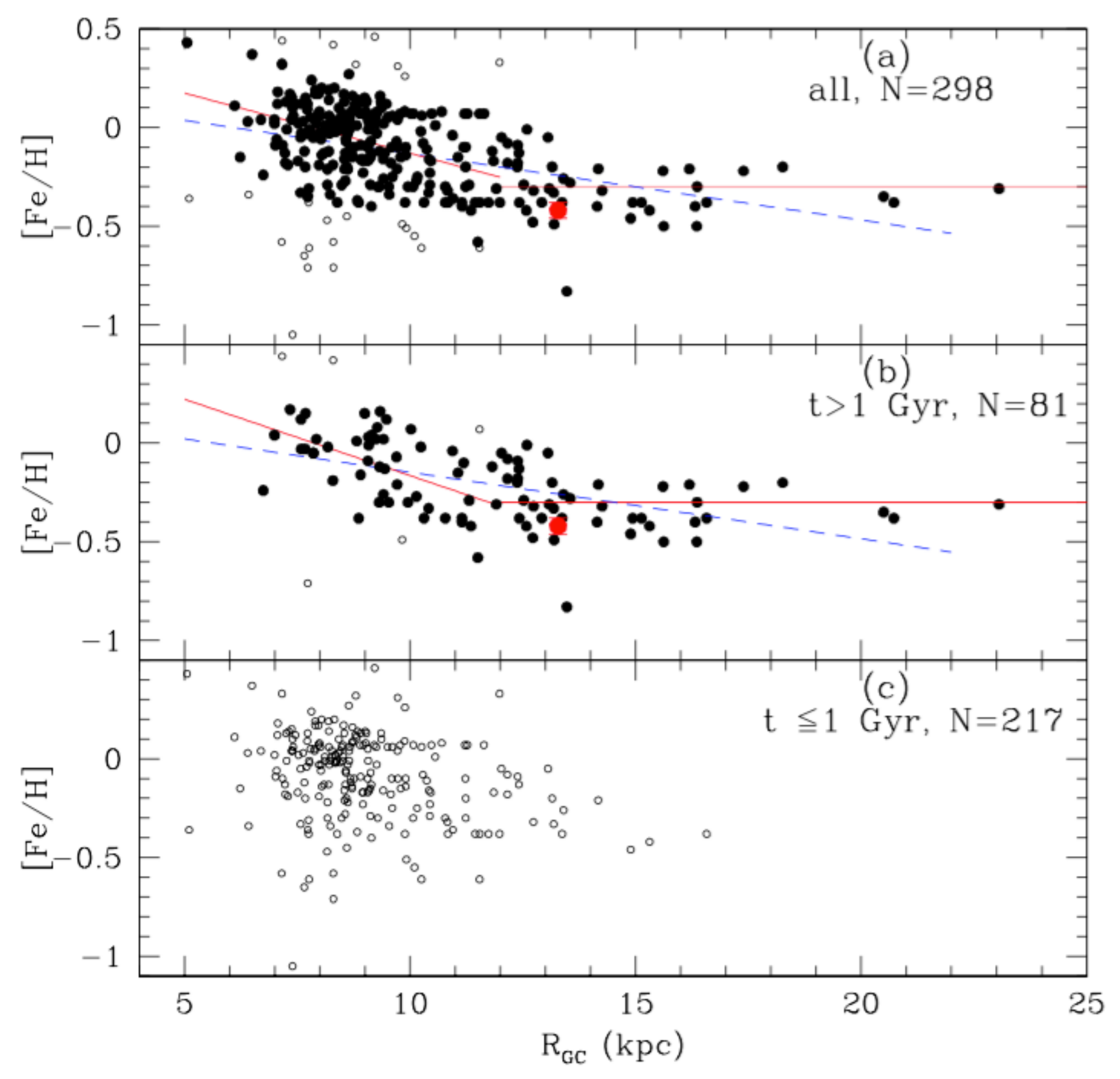}
\caption{Galactocentric radial distribution of the OCs in the Milky Way.
Data come from the DAML02 catalog
  for OCs with both Galactocentric radius ($R_\mathrm{GC}$) and metallicity ([Fe/H]) values.
Panel (a) is for all the 298 OCs, panel (b) is only for the 81 old (age $> 1$ Gyr) OCs, 
  and panel (c) is for the young (age $\le 1$ Gyr) OCs.
Blue dashed lines are the least square fittings for all the data
  for the whole radial range.
Panels (a) and (b) show that 
  OCs in the outer region of $R > 12$ kpc might show almost constant 
  metallicity value ([Fe/H] $\sim -0.3$ dex) regardless of the radius,
  while the red solid lines for the inner region of $R < 12$ kpc
  are the least square fittings only for the OCs remained after 2$\sigma$ clipping 
  (N=231 for panel (a) and N=49 for panel (b)).
Small open circles at $R < 12$ kpc in (a) and (b) 
  denote the OCs excluded during the 2$\sigma$ clipping process,
  and the large red dot with error bars represents the position of Ruprecht 6
  with parameters obtained in this study.
}
\label{fig:Rgc}
\end{figure}

For the OCs in the DAML02 catalog, we have calculated the 
  Galactocentric distances using the distance estimates, Galactic coordinates 
  in the catalog and the equation : 
  $
  R_{\mathrm{GC}} = \sqrt{ [d ~{\rm cos}(b) ~{\rm cos}(l) - R_0]^2 + d^2 ~{\rm cos}^2(b) ~{\rm sin}^2(l) +
   d^2 ~{\rm sin}^2(b)}
  $, where $d$ is the heliocentric distance to the cluster, $l$ and $b$ are the Galactic 
  longitude and Galactic latitude, respectively, of the cluster, $R_0$ is the distance of the Sun
  from the Galactic center ($8.5$ kpc is used in this study).
Out of 2167 OCs listed in the DAML02 catalog,
  only 298 clusters have both of metallicity and distance estimates,
  which are shown in Figure~\ref{fig:Rgc}
  as a function of the Galactocentric distance, $R_\mathrm{GC}$.

We have performed the least square fitting for all the 298 OCs
  with both metallicity and distance estimates, and obtained 
  [Fe/H] $ =(-0.034\pm0.005) R_{\mathrm{GC}} + (0.204\pm0.053)$ with rms = 0.229
  for all the radial range of the Galaxy
  (shown as blue dashed line in Figure~\ref{fig:Rgc} (a)).
Assuming the outer ($R>12$ kpc) clusters might follow constant ([Fe/H] $\sim -0.3$ dex)
  metallicity trend,
  we also made the least square fitting only for the inner 
  ($R_{\mathrm{GC}} < 12$ kpc) OCs
  with 2$\sigma$ clipping, which returns 
  [Fe/H] $=(-0.061\pm0.008) R_{\mathrm{GC}} + (0.479\pm0.071)$ with rms = 0.155
  (N = 231) (shown as red solid line in Figure~\ref{fig:Rgc} (a)).

Taking into account only the old OCs with age $> 1$ Gyr,
  we plotted the same plot in Figure~\ref{fig:Rgc} (b).
The 81 objects in the whole radial range give the relation of
  [Fe/H] $=(-0.034\pm0.007) R_{\mathrm{GC}} + (0.190\pm0.080)$ with rms = 0.201
  (blue dashed line in Figure~\ref{fig:Rgc} (b)), and
  the 49 clusters at $R_\mathrm{GC} < 12$ kpc left after the 2$\sigma$ clipping process returns
  [Fe/H] $=(-0.077\pm0.017) R_{\mathrm{GC}} + (0.609\pm0.161)$ with rms = 0.152
  (red solid line in Figure~\ref{fig:Rgc} (b)).
This result is in very good agreement with those obtained recently by \citet{ryu11} and
  \citet{andreu11}, while the number of OCs used in this study is smaller than
  those of the previous studies which resulted from introduction of the process of extracting
  only the old and good (by 2$\sigma$ clipping) clusters for the analysis.

Whether we adopt the single relation for the whole radial range
  or the dual relation broken at $R_\mathrm{GC} \sim 12$ kpc,
  Ruprecht 6 seems to conform either of the radial metallicity trends
  at its location of $R_{\mathrm{GC}} = 13.28\pm 0.54$ kpc.

The metallicity estimates from photometric indices are less reliable
  than those from spectroscopic observations.
$(U-B)$ colors for late-F to early-K-type stars,
  Washington and DDO photometric systems for G and K-type giants 
  can give some reliable values.
However, while the DAML02 catalog gives detailed references for proper motions
  and radial velocities, the sources for the metallicity estimates are not listed.
Among the 81 objects in Figure~\ref{fig:Rgc} (b),
  there are 18 objects for which $N \ge 10$ stars are used to determine the metallicity,
  40 objects for which $1 \le N \le 9$ stars are used, and
  no information is given for the other 23 objects.
Investigating the sources of the metallicity estimates again is beyond the scope
  of this paper, but compilation and classification of these estimates
  using the criteria of the methods to find abundances and number of stars used
  will be a good approach to obtain more reliable 
  Galactocentric radial metallicity distribution of the OCs in the Milky Way.

\begin{table*}[t!]
\caption{Slope and intercept values in the equation of
  $<[Fe/H]>=a \times R_{\mathrm{GC}} + b$\label{tab:slope}}
\centering
\setlength{\tabcolsep}{1.2mm}
\begin{tabular}{lccccr}
\toprule
Reference & Radial Coverage & $a$ & $b$ & RMS & N \\
\midrule
This study & All        & $-0.034\pm0.007$ & $0.190\pm0.080$ & 0.201  &81\\
This study & $R<12$ kpc & $-0.077\pm0.017$ & $0.609\pm0.161$ & 0.152 & 49 \\
This study & $R>12$ kpc & $ 0            $ & $-0.3 $         & --    & 27 \\
\citet{ryu11}   & $R<12$ kpc & $-0.076\pm0.013$ & $0.600\pm0.116$ & 0.029 & 186 \\
\citet{ryu11}   & $R>12$ kpc & $ 0            $ & $-0.3         $ &       & 186 \\
\citet{andreu11}& All        & $-0.04         $ & --              & --    &177 \\
\citet{andreu11}& $R<12$ kpc & $-0.07         $ & --         & --    &177 \\
\citet{andreu11}& $R>12$ kpc & $ 0            $ & $-0.35$         & --    &177 \\
\citet{kim06}   & All        & $-0.067\pm0.009$ & --              & --    & 53 \\
\citet{kim05}   & All        & $-0.064\pm0.009$ & --              & --    & 51 \\
\citet{kim05}   & --         & $-0.066\pm0.019$ & --              & --    & $\dagger$ \\
\citet{friel02} & All        & $-0.059\pm0.010$ & --              & --    & 39 \\
\citet{piatt95} & All        & $-0.07 \pm0.01 $ & --              & 0.13  & 63 \\
\bottomrule
\end{tabular}
\small{\\
Note.-- Unit of the slope $a$ ($=\Delta$[Fe/H]$/ \Delta R_{\mathrm{GC}}$) is dex kpc$^{-1}$. \\
$\dagger$ Mean value of the nine literature values.
}
\end{table*}

\section{Summary\label{sec:sum}}
We derived the physical parameters of the small-size and poorly studied
  old OC Ruprecht 6
  using the $BVI$ optical photometry data.
The main results obtained in this study are :

$\bullet$
The color-magnitude diagrams of Ruprecht 6 show clear MS stars.
The MS turn-off point is at $V \approx 18.45$ mag and $B-V \approx 0.85$ mag.

$\bullet$
Three RC stars are at $V = 16.00$ mag, $I = 14.41$ mag and $B-V = 1.35$ mag.
The mean $K_s$-band magnitude of RC stars ($K_s=12.39 \pm 0.21$ mag)
  and the known absolute magnitude of the RC stars ($M_{K_S} = -1.595 \pm 0.025$ mag), 
  the distance modulus to Ruprecht 6 is obtained to be
  $(m-M)_0 = 13.84 \pm 0.21$ mag ($d=5.86 \pm 0.60$ kpc).

$\bullet$
From the $(J-K_s)$ and $(B-V)$ colors of the RC stars,
  comparison of the $(B-V)$ and $(V-I)$ colors of the bright stars in Ruprecht 6
  with those of the intrinsic colors of dwarf and giant stars,
  and the PARSEC isochrone fittings, 
  we derive the reddening values of 
  $E(B-V) = 0.42$ mag and $E(V-I) = 0.60$ mag.

$\bullet$
Using the PARSEC isochrone fittings onto the color-magnitude diagrams,
  we obtain the age and metallicity values to be :
  log (t) $=9.50 \pm 0.10$ (t $=3.16 \pm 0.82$ Gyr)
  and [Fe/H] $= -0.42 \pm 0.04$ dex.

$\bullet$
For the old (age $> 1$ Gyr) OCs of DAML02 catalog,
  we obtain the Galactocentric radial metallicity relations of either
  (i) a single relation of 
  [Fe/H] $=(-0.034\pm0.007) R_{\mathrm{GC}} + (0.190\pm0.080)$ (rms = 0.201)
  or (ii) dual relation of 
  [Fe/H] $=(-0.077\pm0.017) R_{\mathrm{GC}} + (0.609\pm0.161)$ (rms = 0.152) 
  at $R_\mathrm{GC} < 12$ kpc and constant ([Fe/H] $\sim -0.3$ dex) value
  at $R_\mathrm{GC} > 12$ kpc.

\acknowledgments
We thank the anonymous referee and the Scientific Editor for the thorough review and helpful comments
  that helped to improve the manuscript.
The participation of I. H. and S. K. in this project was made possible
  by UST Research Internship for Undergraduates grant in 2016 July.
Based on observations at Cerro Tololo Inter-American Observatory, 
  National Optical Astronomy Observatory (NOAO Prop. ID 2010B-0178, PI Sang Chul KIM),
  which is operated by the Association of Universities for Research in Astronomy (AURA)
  under a cooperative agreement with the National Science Foundation.
This publication makes use of data products from the Two Micron All Sky Survey, 
  which is a joint project of the University of Massachusetts and 
  the Infrared Processing and Analysis Center/California Institute of Technology, 
  funded by the National Aeronautics and Space Administration and 
  the National Science Foundation.

\appendix


\end{document}